\documentclass[%
 reprint,
 amsmath,amssymb,
 aps,
prb,
]{revtex4-2}

\usepackage{graphicx}
\usepackage{dcolumn}
\usepackage{bm}
\usepackage{color}  
\usepackage{float}  

\begin{document}

\preprint{v:0.92~~~~~12.07.22}

\title{The magnetic dynamics of NiPS$_3$}

\author{A. R. Wildes}
 \email{wildes@ill.fr}
\affiliation{Institut Laue-Langevin, 71 avenue des Martyrs CS 20156, 38042 Grenoble Cedex 9, France}

\author{J. R. Stewart}
\affiliation{ISIS Pulsed Neutron and Muon Source, STFC Rutherford Appleton Laboratory, Harwell Campus, Didcot, OX11 0QX, UK}

\author{M. D. Le}
\affiliation{ISIS Pulsed Neutron and Muon Source, STFC Rutherford Appleton Laboratory, Harwell Campus, Didcot, OX11 0QX, UK}

\author{R. A. Ewings}
\affiliation{ISIS Pulsed Neutron and Muon Source, STFC Rutherford Appleton Laboratory, Harwell Campus, Didcot, OX11 0QX, UK}

\author{K. C. Rule}
\affiliation{Australian Nuclear Science and Technology Organisation, Locked Bag 2001, Kirrawee DC NSW 2232, Australia}

\author{G. Deng}
\affiliation{Australian Nuclear Science and Technology Organisation, Locked Bag 2001, Kirrawee DC NSW 2232, Australia}

\author{K. Anand}
\affiliation{Institut Laue-Langevin, 71 avenue des Martyrs CS 20156, 38042 Grenoble Cedex 9, France}
\affiliation{Department of Earth and Environmental Sciences, Ludwig-Maximilians-Universit{\"a}t M{\"u}nchen, Theresienstrasse 41, 80333 Munich, Germany}
\affiliation{Lehrstuhl f{\"u}r Funktionelle Materialien, Physik Department, Technische Universit{\"a}t M{\"u}nchen, 85748 Garching, Germany}
\affiliation{Institut de Physique de Rennes, UMR UR1 - CNRS 6251, 35042 Rennes Cedex, France}

\date{\today}

\begin{abstract}
Neutron spectroscopy measurements have been performed on single crystals of the antiferromagnetic van der Waals compound NiPS$_3$.   Linear spin wave theory using a Heisenberg Hamiltonian with single-ion anisotropies has been applied to determine the magnetic exchange parameters and the nature of the anisotropy.  The analysis reveals that NiPS$_3$ is less two-dimensional than its sister compounds, with a relatively large ferromagnetic exchange of $J^{\prime} = -0.3$ meV between the layered \emph{ab} planes.  In-plane magnetic exchange interactions up to the third nearest-neighbour were required to fit the data.  The nearest-neighbour exchange was ferromagnetic with $J_1 = -2.6$ meV, the second neighbour was antiferromagnetic and small with $J_2 = 0.2$ meV, and the dominant antiferromagnetic third neighbour exchange was $J_3 = 13.5$ meV.  The anisotropy was shown to be largely XY-like with a small uniaxial component, leading to the appearance of two low-energy spin wave modes in the spin wave spectrum at the Brillouin zone centre.  The analysis could reproduce the spin wave energies, however there are discrepancies with the calculated neutron intensities hinting at more exotic phenomena.
\end{abstract}

\maketitle


\section{Introduction}

NiPS$_3$ belongs to a family of magnetic layered van der Waals compounds.  The family has been investigated relatively intensely since the late 1960s.  Initial investigations focused on their low-dimensional magnetic behaviour \cite{Brec}.  They then attracted considerable attention for the ability to intercalate other species into the host \cite{Grasso}.  More recently, the discovery of graphene has led to renewed interest in the family as they can be delaminated to monolayer thickness while retaining their magnetism \cite{Park}.  

The family members consist of a magnetic transition metal with a 2+ ionization state, in particular Mn, Fe, Co and Ni, and the sulfur can be substituted for selenium.  The crystal structures for all the thiophosphate compounds share the same $C~2/m$ space group \cite{Ouvrard85}, with the transition metal atoms forming a planar honeycomb lattice in the \emph{ab} planes.  The metal atoms are enclosed in octahedra formed by the sulfur atoms and  there is a phosphorus doublet at the centre of the honeycomb hexagons, with the ensemble forming layers that stack along the {\bf{c}} axis with weak van der Waals bonding.   All the compounds are antiferromagnetic, although they have different magnetic structures and anisotropies \cite{Brec}.

\begin{figure}
  \includegraphics[width=3.5in]{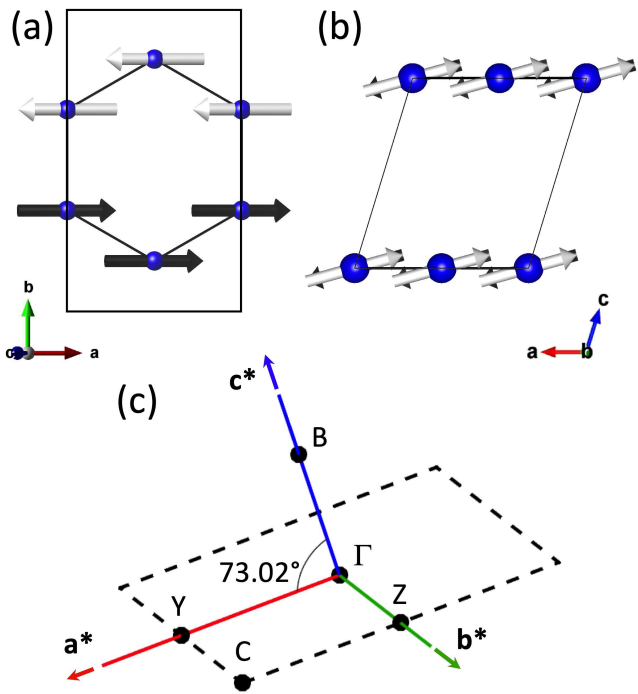}
  \caption{\label{fig:MSBZ} The magnetic structure of NiPS$_3$ \cite{Wildes15}, viewed along the (a) ${\bf{c^*}}$ axis, and (b) {\bf{b}} axis respectively. The figure was created using the VESTA software package \cite{VESTA}.  (c) the Brillouin zone for the magnetic structure of NiPS$_3$.  Selected high-symmetry points are indicated, corresponding to the magnetic space group $P_C2_1/m$ (\#11.57) as defined in the Bilbao Crystallographic Library \cite{Bilbao1,Bilbao2,Bilbao3}.}	
\end{figure}

NiPS$_3$ magnetically orders below $T_N \sim 160$ K with the magnetic structure shown in figure \ref{fig:MSBZ} \cite{Wildes15} along with the corresponding Brillouin zone.   The magnetic moments are collinear, lying mostly along the {\bf{a}} axis with a small component normal to the \emph{ab} planes.  The structure consists of ferromagnetic zig-zag chains parallel to the {\bf{a}} axis that are antiferromagnetically coupled along the {\bf{b}} axis, resulting in the magnetic propagation vector {\bf{k}}$_\textrm{M} = \left[0 1 0\right]$.  Its paramagnetic susceptibility is isotropic \cite{Wildes15}, suggesting that the magnetic Hamiltonian is Heisenberg-like.  The Mermin-Wagner theorem states that a Heisenberg magnet cannot form long-ranged order in two dimensional materials at finite temperatures, however NiPS$_3$ has the highest N{\'e}el temperature of all the members in the thiophosphate family suggesting that interplanar magnetic interactions may be more important than its sister compounds.  This is also suggested by recent Raman spectroscopy measurements of NiPS$_3$ showing that the compound retains its long-ranged magnetic structure when thinned to two layers, but that the order is lost when thinned to a monolayer \cite{Kim}.  

Within the flurry of recent activity concerning the family, there has been particular interest in NiPS$_3$.  It has the smallest spin of the family with $S = 1$, hence may host quantum effects, exotic magneto-lattice and magneto-electron behaviour \cite{Belvin, Hwangbo, Kang}.  Detailed knowledge of the spin dynamics is required to understand the magnetic properties of the compound.  A variety of high-resolution spectroscopy methods have been used to probe the dynamics of NiPS$_3$, including recent efforts with light \cite{Kim, Belvin}, optical pump-probe methods \cite{Afanasiev} and electron spin resonance \cite{Mehlawat}.  These methods give information on the dynamics at the Brillouin zone centre.  Neutron spectroscopy is the method of choice to probe the dynamics throughout the Brillouin zone, and it has been used to measure the spin waves from a sample of powdered NiPS$_3$ \cite{Lancon18}.  The study concluded that the spin dynamics could be modelled using a Heisenberg Hamiltonian with a uniaxial single-ion anisotropy.  The nearest-neighbour exchange  was found to be ferromagnetic, and the antiferromagnetic structure was stabilised by third in-plane nearest neighbour exchange which was shown to be by far the dominant term in the Hamiltonian.  The single-ion anisotropy gave rise to an energy gap at the Brillouin zone centre, and would further stabilise the long-ranged magnetic order.

There were a number of ambiguities and assumptions present in the powder data and their analysis.  The modelling assumed that the compound was purely two-dimensional, with no exchange between the \emph{ab} planes.  The analysis found that the low energy ($\Delta E \le 10$ meV) spin waves had almost equal energies at the Brillouin zone centre at $\Gamma$ and the corners at C.  This is ambiguous in the analysis of powder data as certain $\Gamma$ and C points have approximately the same momentum magnitude, $Q$.  Crystals of the compounds are known to twin with 120$^\circ$ rotations of the \emph{ab} planes about the ${\bf{c^*}}$ axis \cite{Matsumoto}, resulting in the scattering from magnetic $\Gamma$ points being potentially superimposed onto the C points.  Finally, the magnitude of a spin wave gap at the $\Gamma$ point was not clearly determined with the best estimate arriving at $\Delta E \sim 7$ meV.  This is somewhat at odds with more recent measurements, some of which show that there may be two gaps in the low-energy modes at the $\Gamma$ point \cite{Afanasiev, Mehlawat}.  Two gaps would be consistent with an easy-plane anisotropy \cite{Kim}, contrasting with the analysis of the neutron powder data that used a uniaxial anisotropy \cite{Lancon18}.

This article describes a neutron spectroscopy study of the spin dynamics of single crystals of NiPS$_3$ to resolve the ambiguities in the analysis of the powder data.  The data were measured using neutron three-axis and direct-geometry time-of-flight spectrometers, and they were analysed by linear spin wave theory using the SpinW software package \cite{SpinW}.  The discussion reconciles the neutron data with measurements of the spin wave energies at the $\Gamma$ point using other experimental techniques and outlines limitations in using a simple Heisenberg Hamiltonian with straight-forward single-ion anisotropies for the data analysis.

\section{Experiment}
\subsection{Sample preparation}
Crystals of NiPS$_3$ were prepared using a vapour transport method that has previously been described in detail \cite{Wildes15}.  A total of 8 of the largest crystals with a total mass of $\approx 0.5$ g were chosen for the neutron experiments, being platelets with typical dimensions $\gtrsim 5 \times 5 \times 0.2$ mm$^3$.

This family of compounds is known to be prone to characteristic twinning involving the rotation of the {\bf{a}} and {\bf{b}} axes by $120^\circ$ about the ${\bf{c^*}}$ axis \cite{Murayama}, which is normal to the platelets.  The twinning results in diffraction patterns that appear to have 6-fold symmetry.  The degree of twinning in each of the crystals was determined by comparing the integrated intensities of neutron diffraction peaks that do not have something approaching 6-fold symmetry, such as $\{0 2 2\}$ or, in the magnetically ordered state, $\{0 1 0\}$, for the three $120^\circ$ rotations.

The magnetic properties of NiPS$_3$ are affected if crystals are glued directly to supports in preparation for a measurement \cite{Wildes15}, possibly due to strong magnetostriction or intercalation.  Consequently, all the crystals were wrapped in aluminium foil prior to mounting, and glue was never applied directly to a crystal.

\subsection{Three-axis spectrometry}
Neutron three-axis spectrometry was used to study the low energy dispersion from the best crystal.  The crystal had dimensions $\sim 10 \times 10 \times 0.3$ mm$^3$ and a domain ratio of 0.58:0.26:0.16.  The foil-wrapped sample was glued  to an appropriate support using a small amount of GE varnish.  Two supports were used: one to access the $0 k l$ scattering plane, and one to access $h k 0$, defined with respect to the dominant domain.  The sample was simply removed from its aluminium envelope between experiments, then wrapped in a new envelope before being realigned and mounted on the new support.
  
Two three-axis spectrometers were used. 

The IN8 instrument at the Institut Laue-Langevin, France, was used to study the scattering in the $0 k l$ plane \cite{Wildes_IN8_Jun19b}.   The instrument was configured with a pyrolytic graphite (PG) 002 monochromator and analyser.  The wavenumber for the analyser was fixed to $k_f = 2.662$ {\AA}$^{-1}$ for the experiment.  Higher order contamination was suppressed using a PG filter before the analyser.   Both monochromator and analyser were vertically focussed and horizontally flat, and $40^\prime$ collimators were used before and after each of them.  A liquid helium cryostat was used to cool the sample to 1.8 K.

The Taipan instrument at the Australian Nuclear Science and Technology Organisation, Australia, was used to study the scattering in the $h k 0$ plane.  The instrument was configured in the same manner as IN8, with the exceptions that $40^\prime$ collimators were only used before and after the sample and that temperature control down to 5 K was achieved using a closed-cycle refrigerator.

The resolution on both instruments was calibrated by mapping the elastic scattering from a nuclear Bragg peak, and by measuring the energy width of the incoherent scattering at an equivalent position in $Q$.  The $2 0 0$ peak from single crystals of MnPS$_3$ and NiPS$_3$ were used to calibrate IN8 \cite{Wildes_IN8_Jun19} and Taipan respectively.  The measured energy full-width half-maximum of the elastic line was 0.79(3) meV on IN8 and 0.82(3) meV on Taipan. The scattering was subsequently compared to calculations using the Popovici method \cite{Popovici} in the RESCAL5 library \cite{Rescal} for MATLAB\textsuperscript{\textregistered}.  The comparison was satisfactory using the nominal instrument parameters, and a sample mosaic of $25^\prime$.

\subsection{Time-of-flight spectrometry}
Time-of-flight spectrometry experiments were conducted using the MERLIN direct geometry spectrometer at the ISIS spallation neutron source, UK \cite{MERLIN}.  These measurements \cite{Wildes_MERLIN_Oct20} gave an overview of the dynamic structure factor over the entire Brillouin zone.  All eight crystals were combined and co-aligned for the experiment.  The crystals were fixed to aluminium plates using a small amount of GE vanish and some Fomblin\textsuperscript{\textregistered} oil.  The ${\bf{c^*}}$ axis was normal to the plates and the samples were co-aligned such that the $0 k l$ plane was horizontal, defined by the majority domain for the individual crystals.  The plates were then mounted on a common ``toaster rack" aluminium block.  Temperature control to 5 K was achieved using a cryorefrigerator.

The scattering was measured for a series of $1^\circ$ rotations about the normal to the $0 k l$ plane.  A total of 84 rotations were measured.  The scattering could then be combined into a 4-dimensional volume and cut or sliced along desired reciprocal space directions using the HORACE software \cite{Horace}.  MERLIN can profit from rep-rate multiplication, whereby the scattering from multiple incident neutron energies can be measured in a single neutron pulse \cite{Russina,Nakamura}.  Data with $E_i = 16.9, 30.4$ and 70 meV were simultaneously collected.  The resolution for the three incident energies  had a full-width half maximum at $\Delta E = 0$ of 0.5, 1.1 and 3.4 meV respectively.

\section{Results}
\subsection{Neutron three-axis spectroscopy, and the low energy spectra}
\begin{figure}
  \includegraphics[width=3.5in]{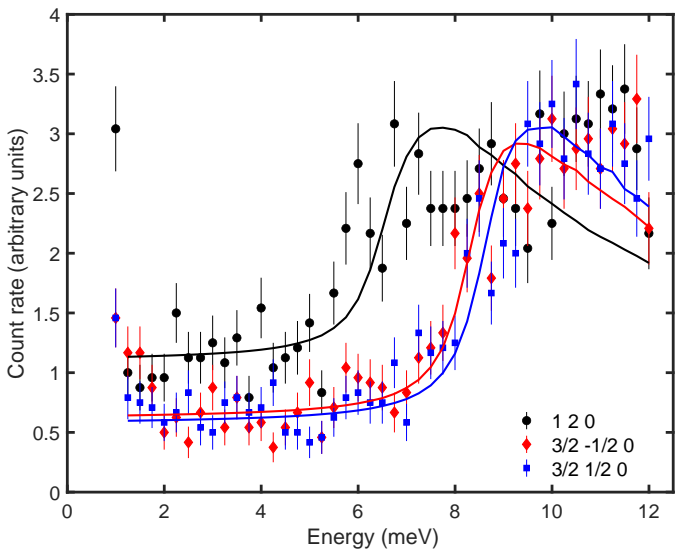}
  \caption{\label{fig:ConstQ} Neutron three-axis spectroscopy data measured using Taipan at {\bf{Q}} = 120, {\bf{Q}} $= \frac{3}{2} \frac{\overline{1}}{2} 0$, and {\bf{Q}} $= \frac{3}{2} \frac{1}{2} 0$.  Corresponding fits to the data are also shown and are discussed in the text.}	
\end{figure}

Three-axis scans show details that were not apparent in the analysis of the data from powdered samples \cite{Lancon18}.  Figure \ref{fig:ConstQ} shows energy scans at ${\bf{Q}} = 1 2 0$, corresponding to $\Gamma$ in figure \ref{fig:MSBZ}c, and at ${\bf{Q}} = \frac{3}{2} \frac{1}{2} 0$ and $\frac{3}{2} \frac{\overline{1}}{2} 0$ corresponding to C points in Brillouin zone corners.  The data at the $\Gamma$ point shows a clear increase at $\Delta E \sim 6$ meV, while a similar increase is seen at $\Delta E \sim 9$ meV for the data at the C points.  These features were expected and correspond to gapped modes observed in the powder data.  The gap at the $\Gamma$ point is consistent with terahertz spectroscopy \cite{Belvin} and optical pump-probe \cite{Afanasiev} measurements that showed features at 5.3 meV  and 3.8 meV respectively.  The intensity does not decrease at larger energy transfer as the line shape is highly asymmetric due to the convolution of the instrument resolution with the very steep spin wave dispersion.

\begin{figure}
  \includegraphics[width=3.5in]{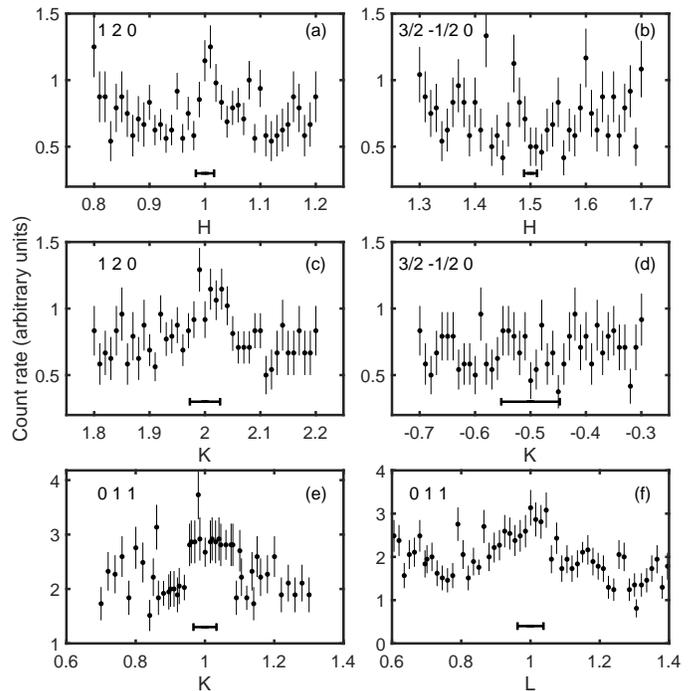}
  \caption{\label{fig:E_2meV} Neutron three-axis scans performed at an energy transfer of $\Delta E~=~2$ meV along different trajectories in reciprocal space.  (a) and (c) respectively show scans parallel to the ${\bf{a^*}}$ and ${\bf{b^*}}$ axes through the Brillouin zone centre at ${\bf{Q}}~=~120$.  (b) and (d) respectively show scans  parallel to the ${\bf{a^*}}$ and ${\bf{b^*}}$ axes through the Brillouin zone corner at ${\bf{Q}} =\frac{3}{2}\frac{\overline{1}}{2}0$.  (e) and (f) respectively show scans  parallel to the ${\bf{b^*}}$ and ${\bf{c^*}}$ axes through the Brillouin zone centre at 0~1~1.  The data in (a)-(d) were measured using Taipan and the data in (e)-(f) were measured using IN8.  Estimates for the resolution-limited width of an infinitely-steep excitation are shown for each scan as horizontal bars.}	
\end{figure}

Close inspection at smaller energy transfers shows that the count rates for $\Delta E \le 5$ meV are systematically larger at $\Gamma$ than at either of the C points, hinting at extra spectral weight.  This is due to a second, lower energy spin wave mode.  Figure \ref{fig:E_2meV} shows constant energy cuts at $\Delta E = 2$ meV, which is approximately the lowest energy transfer where data can be confidently taken to be clean from contamination due to the elastic scattering.  Scans along $h 2 0$ and $1 k 0$  are shown in figure \ref{fig:E_2meV} (a) and (c) respectively and a weak, but clear and consistent, peak is observed at the $1 2 0$ Brillouin zone centre in each.  Similar scans through the C points did not consistently have such a feature, as is observed for the data at ${\bf{Q}} = \frac{3}{2} \frac{\overline{1}}{2} 0$ in figure figure \ref{fig:E_2meV} (b) and (d).  The weak mode is only present at the Brillouin zone centres.  

The weak mode is steeply dispersive along ${\bf{a^*}}$ and ${\bf{b^*}}$, which may be expected, however it is also surprisingly dispersive along ${\bf{c^*}}$ considering that NiPS$_3$ was thought to be quasi-two dimensional.  Figure \ref{fig:E_2meV} (e) and (f) show scans along $0 k 1$ and $0 1 l$.  Figure \ref{fig:E_2meV} (c) and \ref{fig:E_2meV} (e) follow the same trajectory through the Brillouin zone and may be compared directly, both showing weak peaks at the respective Brillouin zone centres.  A peak is also observed at $0 1 1$ in figure \ref{fig:E_2meV} (f).  A true two-dimensional layered compound would have no magnetic exchange between the planes.  Any spin waves normal to the planes would be dispersionless, and scans in this direction would show no peaks.  That there is a peak in figure \ref{fig:E_2meV} (f) shows that there is a magnetic exchange between the layered planes in NiPS$_3$ which appears to be considerably larger than that for the sister compounds MnPS$_3$ \cite{Wildes98} and FePS$_3$ \cite{Lancon}.
 
\begin{figure}
  \includegraphics[width=3.5in]{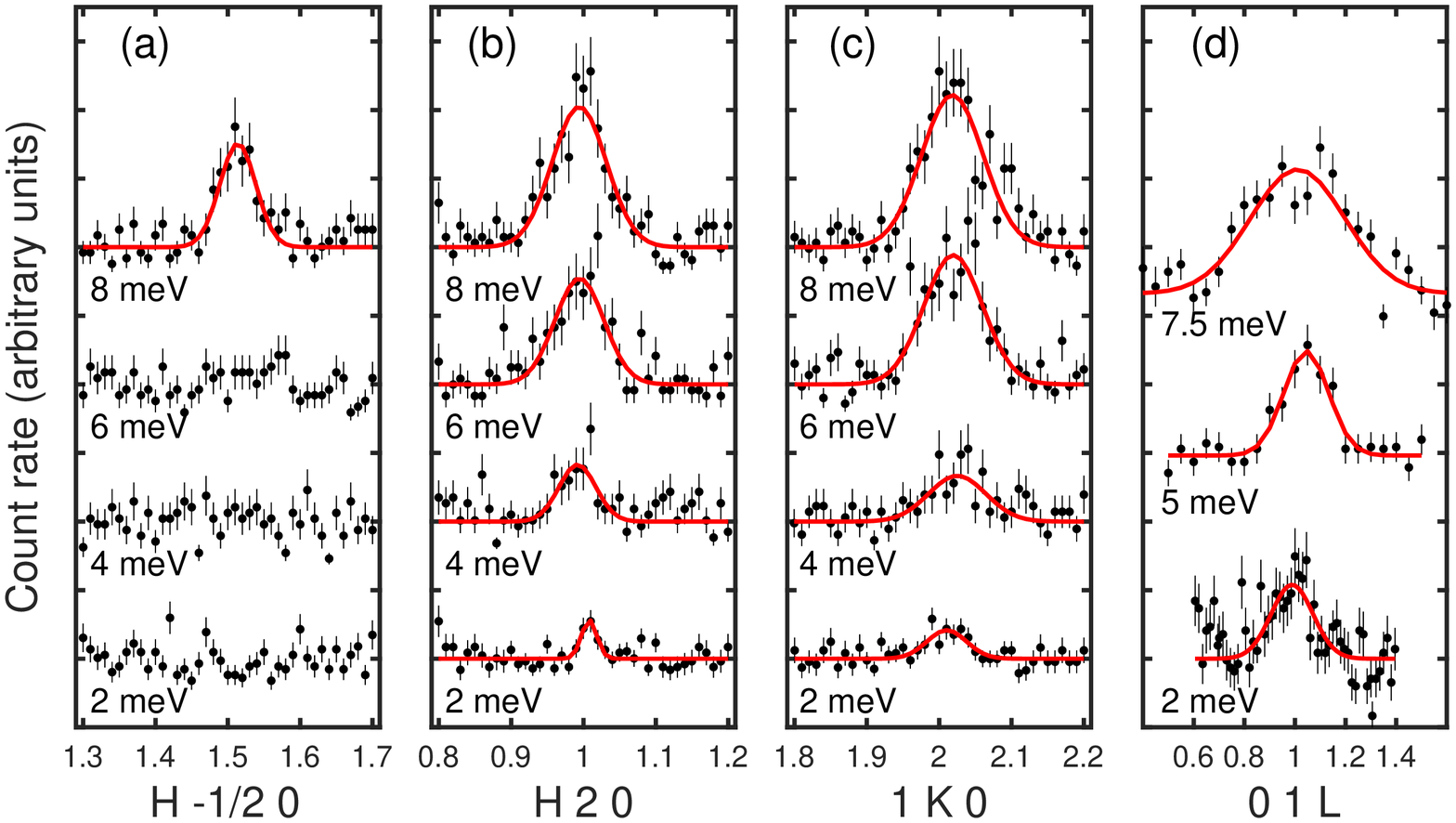}
  \caption{\label{fig:ConstE} Neutron three-axes scans performed at constant energy transfers at different reciprocal lattice points, together with fits of a Gaussian function where appropriate.  Data at equivalent reciprocal lattice points and different energies have been separated vertically for clarity.  (a) Scans parallel to the ${\bf{a^*}}$ axis through the Brillouin zone corner at ${\bf{Q}} = \frac{3}{2}\frac{\overline{1}}{2}0$. (b) and (c) Scans respectively parallel to the ${\bf{a^*}}$ and ${\bf{b^*}}$ axes through the Brillouin zone centre at ${\bf{Q}} =120$. (d) Scans parallel to the ${\bf{c^*}}$ axis through the Brillouin zone centre at ${\bf{Q}} =011$.  The data in (a)-(c) were measured using Taipan and the data in (d) were measured using IN8.}	
\end{figure}

Similar scans at larger energy transfers give an indication of the dispersion of the weak mode.  {\bf{Q}} scans along $h \frac{\overline{1}}{2} 0$, $h 2 0$, $1 k 0$ and $0 1 l$ are shown in figures \ref{fig:ConstE} (a)-(d) respectively.  Figure \ref{fig:ConstE} (a) shows that no clear peaks are observed in the Brillouin zone corner until an energy transfer of $\Delta E \sim 8$ meV, however peaks are present in all the scans at the $1 2 0$ and $0 1 1$ Brillouin zone centres.  The peak intensities are very weak until $\sim 6$ meV, where they increase dramatically.  

\begin{figure}
  \includegraphics[width=3.5in]{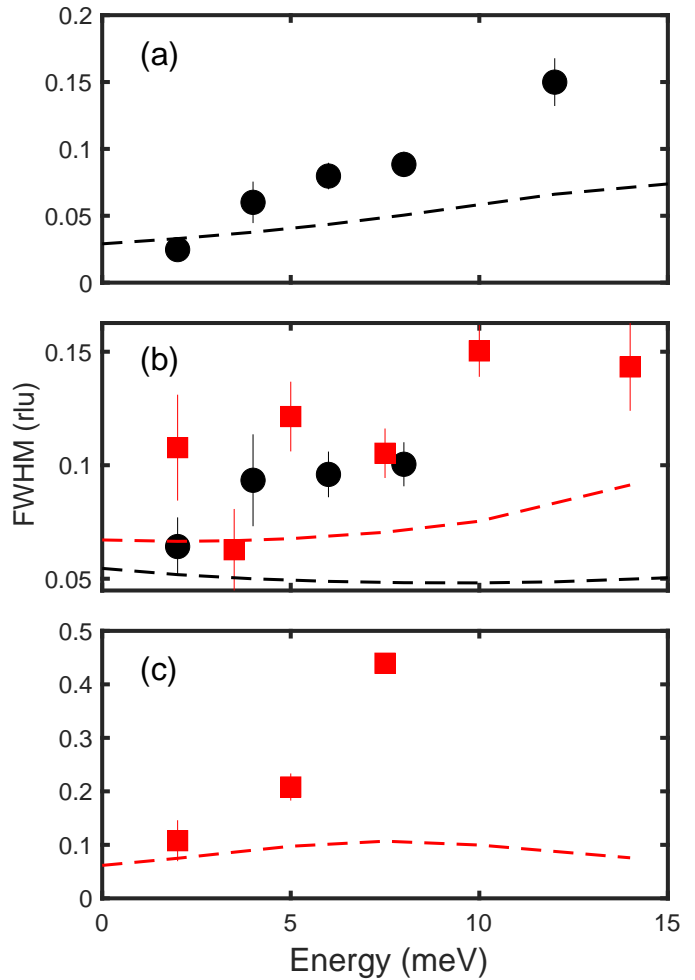}
  \caption{\label{fig:Qwidths} Fitted Gaussian full-widths-at-half-maximum for neutron three-axes scans performed at constant energy transfers.  The scan trajectories are parallel to the (a) ${\bf{a^*}}$, (b) ${\bf{b^*}}$ and (c) ${\bf{c^*}}$ axes respectively.  The black points correspond to measurements from Taipan centred at ${\bf{Q}} =120$, and red points to measurements from IN8 centred at ${\bf{Q}} =011$.  The black and red dashed lines show the calculated widths for the corresponding black and red points if the dispersion was infinitely steep, corresponding to a delta function in {\bf{Q}}}.	
\end{figure}

Dispersive modes would give two peaks in the constant energy cuts shown in figure \ref{fig:ConstE}.  Unfortunately, the instrumental resolution is too coarse to separate the two peaks,  however the fitted widths of a single peak function give an indication of the dispersion.  Gaussian fits to the peaks are shown in the figure, and the  {\bf{Q}} widths in $h$, $k$ and $l$ widths are plotted as a function of energy transfer in figure \ref{fig:Qwidths} (a)-(c).  Data determined from fits around the $1 2 0$ and $0 1 1$ positions are shown as black and red points respectively.  Estimates for the instrumental resolution are also shown as black and red dashed lines, correlated to the data points of the same colour.  The estimates give the expected widths if the mode was a delta function in {\bf{Q}}, corresponding to a spin wave that was infinitely steep.  The data at small energy transfers are close to the resolution estimates, but increase approximately linearly with energy, as expected from linearly dispersive excitations.

The measurements had insufficient resolution to determine whether the lowest energy mode at the $\Gamma$ point was gapped.  A dedicated measurement with a cold neutron spectrometer would be required for this task.  However, it is highly likely to be gapped as the ordered moments are ultimately collinear with a well-defined crystallographic axis, suggesting a non-zero magnetic anisotropy.  Measurements using optical pump-probe \cite{Afanasiev} and electron spin resonance \cite{Mehlawat} techniques have found evidence for a low-energy gap, estimated to be 1.24 meV and 1 meV respectively.

\subsection{Time-of-flight spectroscopy}

\begin{figure}
  \includegraphics[width=3.5in]{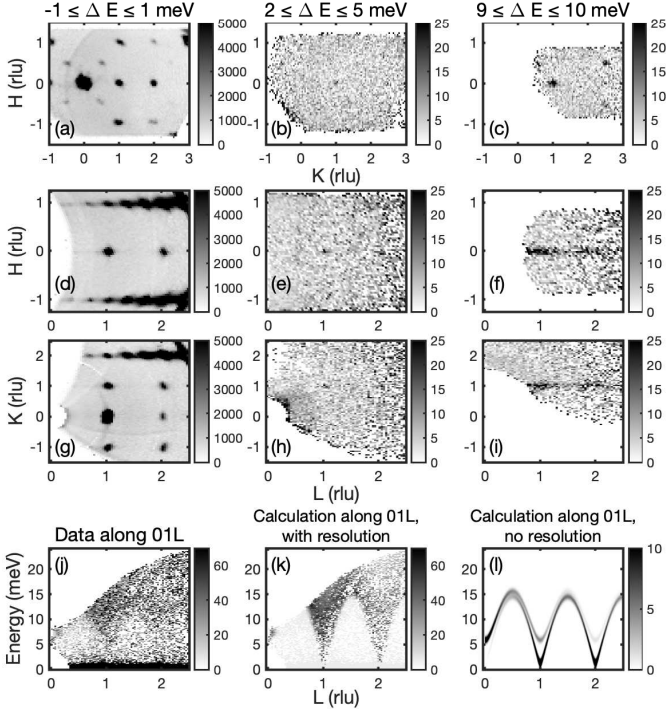}
  \caption{\label{fig:MERLINlo} Neutron spectroscopy measurements from MERLIN, focusing on the low energy modes and the dispersion along ${\bf{c^*}}$. Experimental data at positive and negative $l$ have been combined.  
  (a) - (c) Constant energy slices in $hk\left|1\right|$, with $-0.1 \le \left|l\right| \le 0.1$ rlu. 
  (d) - (f) Constant energy slices in $ h 1 \left|l\right|$, with $-0.1 \le k \le 0.1$ rlu.
  (g) - (i) Constant energy slices in $0 k \left|l\right|$, with $-0.1 \le h \le 0.1$ rlu.
 Data in (a), (d) and (g) are centred at $\Delta E = 0 \pm 1$ meV; in (b), (e) and (h) at $\Delta E = 3.5 \pm 1.5$ meV; and (c), (f) and (i) at  $\Delta E = 10 \pm 1$ meV.
 (j) {\bf{Q}}$,\Delta E$ slice along $0 1 l$, measured with $E_i = 30.4$ meV. 
 (k) Combined SpinW/HORACE calculation of $S_\perp\left({\bf{Q}},\Delta E\right)$ for the same slice shown in (j), accounting for the instrument resolution.  The data have had a small constant added (5 counts) to resemble background. 
 (l) SpinW calculation for the same slices as shown in (j) and (k), without the instrument resolution.}	
\end{figure}

The MERLIN measurements performed with $E_i = 16.9$ meV and 30.4 meV confirm the presence of the low energy mode observed in the three axis data.  Examples of the scattering for a selection of slices through the four-dimensional $\left({\bf{Q}},\Delta E\right)$ data are shown in figure \ref{fig:MERLINlo}.  The scattering was very weak, and statistics were improved by slicing and combining the data along symmetrically identical directions.  In particular, data at the same magnitude of $l$ were combined.  

Figures \ref{fig:MERLINlo}(a), (d) and (g) show the nominally elastic ($\Delta E = 0$) scattering, measured with $E_i = 16.9$ meV, in the $\left(hk1\right)$, $\left(h1l\right)$ and $\left(0kl\right)$ planes respectively.  Nuclear Bragg peaks in the $C \frac{2}{m}$ space group have systematic absences for $h + k \ne 2n$, where $n$ is an integer.  Clear magnetic Bragg peaks, representing $\Gamma$ points in figure \ref{fig:MSBZ}(c), are visible at the expected $\left[010\right]$ propagation vector positions, such as $011$ and $1 2 1$.  Weaker peaks are also visible at the $\left\{\frac{1}{2} \frac{1}{2} 1\right\}$ positions in figures \ref{fig:MERLINlo}(a).  These are due to $0 1 1$ magnetic Bragg peaks from the $120^\circ$ rotational domains.  They are not exactly in the scattering plane as ${\bf{c^*}}$ is not orthogonal to $hk1$ and a $120^\circ$ rotation about this axis will map $0 1 1$ from one of the domains onto a lattice position with $l = \frac{1}{6}$ or $\frac{5}{6}$ in the figure.  However, the peaks are extended along the ${\bf{c^*}}$ axis due to their quasi-two-dimensional nature, as is apparent in figures \ref{fig:MERLINlo}(d) and (g).  Furthermore, the data in figure \ref{fig:MERLINlo}(a) is integrated over $\pm 0.1$ rlu along $l$, and the peaks from the other domains have sufficient intensity that they are observable in the figure.

The low energy spin waves seen in figures \ref{fig:E_2meV} and \ref{fig:ConstE} were also observed on MERLIN.  Figures \ref{fig:MERLINlo}(b), (e) and (h) show constant energy slices with data  integrated over energies from $2 \le \Delta E \le 5$ meV for the same planes and incident energy as shown in figures \ref{fig:MERLINlo}(a), (d) and (g) respectively.  The energy integration limits were chosen to be as large as possible while avoiding contamination from either elastic scattering or contributions from the stronger modes at higher energies.  Very weak intensity is observable at the $\Gamma$ points $0 1 1$ in all the figures, and arguably at $0 1 2$ in figures \ref{fig:MERLINlo}(e) and (h).  The peaks are very sharp in {\bf{Q}} in all three figures, reflecting the dispersion of the spin waves in three dimensions.  No extra intensity is observed at any of the Brillouin zone corners, or at the $\left\{\frac{1}{2} \frac{1}{2} 1\right\}$ positions in figures \ref{fig:MERLINlo}(b), confirming that  the low energy spin waves are only present at the Brillouin zone centres and that the instrumental resolution is sufficient that the slices are free of similar contributions from crystal twins.

Higher energy slices, integrating over $9 \le \Delta E \le 11$ meV, are shown in figures \ref{fig:MERLINlo}(c), (f) and (i).  Figure  \ref{fig:MERLINlo}(c) shows data at this energy that are consistent with those measured using three-axis spectroscopy, with clear intensity spots appearing at the $\Gamma$ point at ${\bf{Q}} =011$ and at the C points at ${\bf{Q}} =\pm\frac{1}{2} \frac{5}{2}  1$.  

The extent of the dispersion along ${\bf{c^*}}$ is apparent in figures \ref{fig:MERLINlo}(f) and (i), where the sharp spots at lower energy have become streaks along $\left(01\l\right)$ due to the appearance of the more intense and larger-gapped mode.  The streaks show that this energy is close to the maximum in the dispersion along this direction.  The dispersion is apparent in figure \ref{fig:MERLINlo}(j), which shows a $\left({\bf{Q}},\Delta E\right)$ slice along $\left(01l\right)$ measured with $E_i = 30.4$ meV neutrons.  The higher-intensity mode is the most clearly visible, with minima at $l =$ 1 and 2 and an energy of $\sim 13$ meV at the $B$ point at $l = \frac{3}{2}$.  Individual modes are not observed as the slice was integrated over $\pm 0.1$ rlu in both $h$ and $k$, and the dispersion in these directions is so steep that the spin waves in these directions are also included in the integration.
 
\begin{figure}
  \includegraphics[width=3.5in]{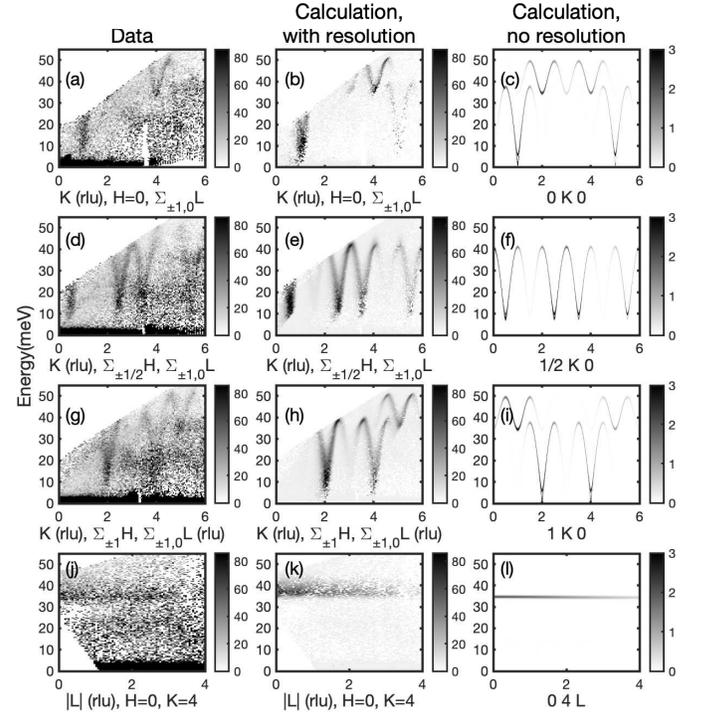}
  \caption{\label{fig:MERLINhi} Neutron spectroscopy {\bf{Q}}$,\Delta E$ slices from MERLIN.  All data were collected with $E_i = 70$ meV. (a) Combined slices from $0 k 0$ and $0 k \left|1\right|$, with corresponding SpinW/HORACE calculations including (b) and without (c) convolution with the instrument resolution. (d) Combined slices from $\left|\frac{1}{2}\right| k 0$ and $\left|\frac{1}{2}\right| k \left|1\right|$, with corresponding SpinW/HORACE calculations including (e) and without (f) convolution with the instrument resolution. (g) Combined slices from $\left|1\right| k 0$ and $\left|1\right| k \left|1\right|$, with corresponding SpinW/HORACE calculations including (h) and without (i) convolution with the instrument resolution. (j) Slice from $0 4 \left|l\right|$, with corresponding SpinW/HORACE calculations including (b) and without (c) convolution with the instrument resolution with corresponding SpinW/HORACE calculations including (k) and without (l) convolution with the instrument resolution.  The SpinW/HORACE plots in (b), (e), (h) and (k) have had a small constant added (5 counts) to resemble background.}	
\end{figure}

$\left({\bf{Q}},\Delta E\right)$ slices through the data taken with $E_i = 70$ meV show the dispersion along high-symmetry directions at higher energies.  Figures \ref{fig:MERLINhi} show representative slices, with (a), (d) and (g) showing slices parallel to ${\bf{b^*}}$ for selected values of $h$ and integer values for $l$ and (j) showing a slice along $\left(04l\right)$.  Qualitative inspection showed that slices in equivalent reduced lattice units were identical, and the data were subsequently combined for better statistics.

Inspection of the data shows the presence of a third mode at the $\Gamma$ point, in addition to the low-energy modes shown in the three-axis measurements and the data in figure \ref{fig:MERLINlo}.  It appears most clearly as a minimum at $\Delta E \sim 34$ meV in a steeply dispersive mode at $k = 4$ in figure \ref{fig:MERLINhi}(a) and at $k = 5$ in  figure \ref{fig:MERLINhi}(g).  A slice along $\left(04l\right)$ reveals that this mode is dispersionless within the resolution of the measurement, as shown in figure \ref{fig:MERLINhi}(l).

On the other hand, while the spectral weight varies as a function of the reduced lattice units, only one clear mode is observed in the slices along the Y$-$C Brillouin zone boundary, as shown in figure \ref{fig:MERLINhi}(d).  

The data indicate that there are at least three spin waves in the compound and that some branches are close to, or potentially exactly, degenerate.  The observation is consistent with linear spin wave theory which predicts four spin waves, one for each magnetic atom in the unit cell, some of which may be degenerate.

\section{Analysis with linear spin wave theory}
Within the resolution of the measurements, the spin dynamics of NiPS$_3$ consists of well-defined spin waves and can be analysed using linear spin-wave theory with an appropriate Hamiltonian.  NiPS$_3$ has four moments in the magnetic unit cell, and in principle this results in the Hamiltonian being expressed as an $8 \times 8$ matrix.  Previous analysis of neutron inelastic scattering from a powdered sample \cite{Lancon18} used a Heisenberg Hamiltonian with a single-ion anisotropy,
\begin{equation}
  \hat{\mathcal{H}}=\frac{1}{2}\sum_{\left< ij \right>}J_{ij}{\bf{S}}_i\cdot{\bf{S}}_j+D\sum_i\left(S^x_i\right)^2,
  \label{eq:Hamiltonian_1}
\end{equation}
where $J_{ij}$ is the exchange interaction between neighbours {i} and {j}, and $D$ is the anisotropy strength.  The previous analysis assumed a uniaxial anisotropy with the $x$ axis collinear to the aligned moment directions.  This simplifies the Hamiltonian, allowing it to be expressed as a $4 \times 4$ matrix which can be analytically diagonalised to give expressions for $S\left({\bf{Q}},\Delta E\right)$.  Many of the spin wave branches become degenerate. Notably, there are only two doubly-degenerate modes at the $\Gamma$ point, one at low and one at high energy.  Also, only one four-fold degenerate mode is present along the Y-C Brillouin zone boundary.

Subsequent experiments using Raman spectroscopy \cite{Kim} suggested that the compound is better described using an XXZ Hamiltonian:
\begin{equation}
  \begin{split}
  \hat{\mathcal{H}}=\frac{1}{2}\sum_{\left< ij \right>}J_{ij} &\left(S^x_iS^x_j + S^y_iS^y_j + \alpha S^z_iS^z_j \right) \\ 
  &+ D^x\sum_i\left(S^{x}_i\right)^2 + D^z\sum_i\left(S^{z}_i\right)^2,
  \label{eq:Hamiltonian_2}
  \end{split}
\end{equation}
with $x$ again parallel to the aligned moment direction and {z} normal to both this direction and {\bf{b}}.  The parameter $\alpha$ permits anisotropy in the magnetic exchange parameters.  This Hamiltonian must be expressed as an $8 \times 8$ matrix which is not easily diagonalised analytically, but can be solved numerically using the SpinW software package \cite{SpinW} written for MATLAB\textsuperscript{\textregistered}.  The program is also compatible with HORACE \cite{Horace} and the combined software can be used to calculate the expected neutron scattering, including a convolution with the instrumental resolution.   The more general Hamiltonian in equation \ref{eq:Hamiltonian_2} lifts the degeneracy of the spin wave branches in a manner consistent with the observation of two low-energy modes at the $\Gamma$ point, and hence equation \ref{eq:Hamiltonian_2} was used to analyse the data.

Manipulation of the parameters shows that the anisotropy must be dominated by an easy-plane single-ion anisotropy largely confined to the \emph{ab} planes, i.e. by a relatively large and positive $D_z$.  The value for $D^x$ must be small, and $\alpha$ must be very close to one.  Departures from these conditions led to large and dramatic discrepancies with the data, in particular with the appearance of multiple modes along the Y-C Brillouin zone boundaries.  A value of $\alpha = 1$ was subsequently adopted to reduce the number of free parameters for all subsequent analysis.

SpinW was not explicitly used to fit the data as they had relatively poor statistics and a complicated background that made fitting two-dimensional data sets problematic.  Instead, data were extracted at given high-symmetry points in the Brillouin zone where peaks due to spin waves were least ambiguous.  The peaks were fitted to determine the spin wave energies, and the energies could then be fitted using analytical expressions.  

The spin wave energies determined from the data are listed in table \ref{tab:DataFit}.  A total of 7 data points were extracted. The low-energy excitations at the $\Gamma$ and C points were extracted from the three-axis data.  The fits included a flat background and a convolution with the instrumental resolution, and examples are shown in figure \ref{fig:ConstQ}.  The spin wave energy at the B point was extracted from the MERLIN data collected at $E_i = 30.4$ meV, shown in figure \ref{fig:MERLINlo}(j), and energies for $\Delta E > 15$ meV were extracted from the data at $E_i = 70$ meV, shown in figure \ref{fig:MERLINhi}.  All the MERLIN data were fitted with Gaussians. Table \ref{tab:DataFit} also lists their corresponding reciprocal and reduced-reciprocal lattice positions.  An additional point was added to account for a gap in the lowest energy mode at $\Gamma$ position.  The energy was estimated to be at $1.1 \pm 0.1$ meV from the average of the measurements using optical pump-probe \cite{Afanasiev} and electron spin resonance \cite{Mehlawat} techniques.  

SpinW is capable of calculating an analytical expression for the spin waves from small unit cells using the MATLAB\textsuperscript{\textregistered} symbolic toolbox, but it was not capable of returning a general expression from the $8 \times 8$ matrix required for NiPS$_3$.  However, it was able to produce expressions for the eigenvalues at the high-symmetry points.  The expressions were calculated in a manner consistent with previous analysis of the transition metal-PS$_3$ compounds \cite{Wildes98, Lancon, Kim}, with magnetic exchange up to the third nearest-neighbour in the $ab$ planes and to the nearest neighbours between the planes.  The spin wave expressions therefore contained only six free parameters: $J_{1..3}$ for the intraplanar exchange between first to third neighbours, $J^{\prime}$ for the interplanar exchange, $D^x$ for a uniaxial anisotropy along the collinear moment axes and $D^z$ for an easy-plane anisotropy, with the spin taken to be $S = 1$.  The expressions are listed in the appendix.

\begin{table}
\caption{Table showing the spin wave energies at specific reciprocal lattice points, extracted from the experimental data.  The data point marked with an asterisk is an average taken from optical pump-probe \cite{Afanasiev} and electron spin resonance \cite{Mehlawat} measurements.  The corresponding eigenvalue attribution is defined with respect to the order in table \ref{tab:Eigenvalues} in the appendix, and the fitted energy for that eigenvalue is also listed.  All energies are in meV.}
\label{tab:DataFit}
\begin{ruledtabular}
\begin{tabular}{ccccc}
{\bf{Q}} & k-vector & Attributed         & Energies from & Fit to   \\
	     &              & eigenvalue       & experiment     & eigenvalue \\
\hline
0 0 0$^\ast$					& $\Gamma$ 	& 1 	& 1.1(1)	& 1.3 \\
 1 2 0    						& $\Gamma$ 	& 2 	& 6.5(2) 	& 5.8 \\
 0 4 0 						& $\Gamma$ 	& 3 	& 35.1(3) 	& 34.7 \\
 $\frac{1}{2} 3 0$				& Y 			& 1 	& 41.1(3) 	& 41.0 \\
 $0 1 \frac{3}{2}$				& B 			& 1 	& 13.8(2) 	& 15.6 \\
 $1 \frac{7}{2} 0$ 				& Z 			& 1 	& 39.4(5) 	& 37.5 \\
 $0 \frac{9}{2} 0$ 				& Z 			& 2 	& 49.0(2) 	& 49.4 \\
 $\frac{3}{2} \frac{\overline{1}}{2} 0$ 	& C 			& 1 	& 8.5(1) 	& 7.7 \\
 \end{tabular}
\end{ruledtabular}
\end{table}

The spin wave energies were then assigned to one of the up-to-four modes calculated at the appropriate wavevector position with the corresponding analytical expression, and the ensemble was then fitted with a Levenberg-Marquardt regression algorithm.  The fitted energies are listed in table \ref{tab:DataFit}, and the output parameters are listed in table \ref{tab:Exchanges}.  The assigned eigenvalues used for the fitting, defined in table \ref{tab:Eigenvalues} in the appendix, are also listed in table \ref{tab:DataFit}.  The choice was somewhat arbitrary, especially at higher energy transfers, as many modes are almost doubly- or, at the Y-point, quadruply-degenerate.  Choosing another mode made little difference to the final fit parameters.   

\begin{table}
\caption{Table showing the fitted exchange parameters from the current data, along with the previously published results from Lan{\c{c}}on \emph{et al.} \cite{Lancon18} and Kim \emph{et al.} \cite{Kim}.  Negative $J$ denotes a ferromagnetic exchange.  $D^x$ and $D^z$ are the magnitudes of the single ion anisotropies along the ${\bf{a}}$ and ${\bf{c^*}}$ axes respectively.  A negative $D$ denotes a uniaxial anisotropy, and a positive value denotes a planar anisotropy.}
\label{tab:Exchanges}
\begin{ruledtabular}
\begin{tabular}{cccc}
 & Reference \cite{Lancon18} & Reference \cite{Kim}  & Current work \\
 \hline
$J_1$ 	& $-3.8$ 	& 3.18 	& $-2.6\left(2\right)$\\
$J_2$ 	& $0.2$ 	& 4.82	& $0.2\left(1\right)$ \\
$J_3$	& $13.8$ 	& 9.08	& $13.5\left(3\right)$\\
$\alpha$	& 1		& 0.66	& 1 \\
$J^\prime$& $-$ 	& $-$	& $-0.32\left(3\right)$ \\
$D^x$	& $-0.3$ 	& $-0.89$ & $-0.010\left(9\right)$\\
$D^z$ 	& 0 		& 2.85 	& $0.21\left(8\right)$\\
 \end{tabular}
\end{ruledtabular}
\end{table}

Having relatively few data points with respect to the number of free parameters and the quasi-degeneracy of many of the modes gave unreasonable uncertainties in the fitted parameters when they were all permitted to be free.  Their stabilities were therefore tested by initially fitting with all the parameters free to find the global solution, and then fixing all but one of them in turn for subsequent fits.  The parameters did not change in these fits, and the resulting uncertainties are listed in table II.

The fitted parameters are shown in table \ref{tab:Exchanges}, alongside the previously-published parameters determined from neutron spectroscopy on a powdered sample \cite{Lancon18} and from modelling the two-magnon spectra from Raman spectroscopy \cite{Kim}.  All values are consistent with the Hamiltonian convention used by SpinW.

\begin{figure}
  \includegraphics[width=3.5in]{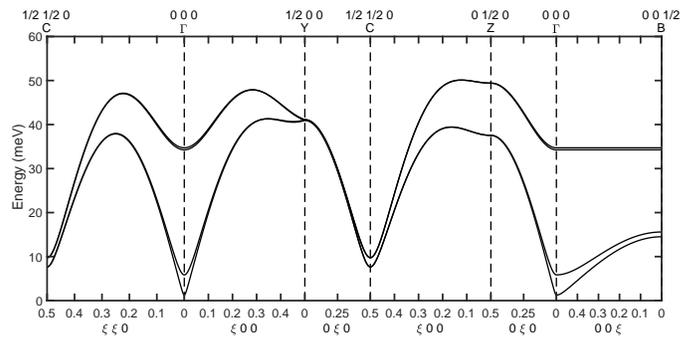}
  \caption{\label{fig:Spag} The spin wave dispersions calculated with the fitted exchange parameters for NiPS$_3$ using equation \ref{eq:Hamiltonian_2} with the parameters listed for the current analysis in table \ref{tab:Exchanges}.}	
\end{figure}

The exchange parameters determined by neutron spectroscopy on powdered and single crystal samples are consistent showing that $J_1$ is ferromagnetic, $J_2$ is much smaller and is antiferromagnetic, and that the magnetic dynamics are dominated by a very large antiferromagnetic third nearest-neighbour interaction, $J_3$.  The magnitude of $J_1$ is smaller in the current analysis than that determined from the powder data, however this is compensated by the inclusion of a relatively large ferromagnetic exchange $J^{\prime}$.  The magnitude of the anisotropy is approximately the same, but the identification of the new low-energy mode at the $\Gamma$ point establishes that it is dominantly co-planar with a very small collinear contribution along the aligned moment direction.  The spin wave energies throughout the Brillouin zone have been calculated using the new set of parameters and is shown in figure \ref{fig:Spag}.

By contrast, the exchange parameters determined from Raman spectroscopy are all approximately the same magnitude and are antiferromagnetic, and the anisotropies are also considerably larger in magnitude.  The spin wave dispersion calculated using these parameters does not match the neutron data, and hence these parameters do not appear to be representative of NiPS$_3$.

The energies determined from using linear spin wave theory with the parameters in table \ref{tab:Exchanges} are listed in table \ref{tab:DataFit}, and they match the data acceptably well.   The parameters were subsequently used with the combined SpinW/HORACE software to calculate the neutron scattering both with and without a convolution with the instrumental resolution.  The calculations are shown in figures \ref{fig:MERLINlo} and \ref{fig:MERLINhi}, and are directly comparable to the experimental data on the corresponding row.  The agreement between measured data and the calculations are mostly satisfactory, confirming that the fitted parameters accurately describe the spin wave dynamics.  However, there are some discrepancies, hinting at limitations in the analysis method to describe the scattering.

The clearest discrepancies are visible when comparing the measured data in figure \ref{fig:MERLINlo}(j) with the corresponding calculations in \ref{fig:MERLINlo}(k) and (l).  The calculations give the lower energy mode significantly more spectral weight than the higher energy mode at $l = 1$ and 2, whereas the neutron data in figures \ref{fig:ConstQ} and \ref{fig:MERLINlo} show the opposite.  According to linear spin wave theory, the lowest energy mode is due to $S^yS^y$ fluctuations while the higher energy mode is due to $S^zS^z$ fluctuations.  This is expected as the relatively large positive value of $D^z$ favours the moments to lie in the $xy$ planes, hence the in-plane fluctuations will cost less energy than those out-of-plane.  The $S^yS^y$ fluctuations give no intensity at ${\bf{Q}} = 010$ as $y \| {\bf{b^*}}$, and thus they have no component perpendicular to the momentum transfer {\bf{Q}}.  However, they become readily visible, and more intense than the $S^zS^z$ fluctuations, with increasing $l$.  That the measured intensities do not match the calculations suggests that the modes are not so simply polarized.

The three-axis data at ${\bf{Q}} =120$, shown in figure \ref{fig:ConstQ}, hint at another discrepancy concerning the spectral weight at the $\Gamma$ point.  The data were fitted with a function to determine the energy gap associated with the larger of the two low-energy modes, and the fit is shown in the figure.  The fit included a convolution with the estimated resolution function, hence the line shape is asymmetric due to the steepness of the dispersion, which was set from inspection of the MERLIN data.  The fit matches the increase of the intensity at $\Delta E \sim 6$ meV, but decreases more rapidly with increasing energy than the data.  The extra intensity in the data at $\Delta E \gtrsim 10$ meV may be due to spurious scattering or contamination from other twinned domains, although an overall view of the experimental results suggest that the latter contribution is minor.  Alternatively, the extra intensity may be the result of more exotic effects, such as quantum fluctuations or hybridization.

Other discrepancies include the fitted value for the fitted dispersion along 00$l$ being slightly steeper than observed, as may be seen by comparing figures \ref{fig:MERLINlo}(j) and (k), and indications of spectral weight at $\Delta E \sim 50$ meV at the ${\bf{Q}} =\frac{1}{2} 4 0$ and $\frac{1}{2} 5 0$ positions, hinted at in \ref{fig:MERLINhi}(d).  The former may be within the uncertainty of the fit, however, this is not the case for the latter.  The calculation in \ref{fig:MERLINhi}(e) shows that some extra intensity at higher energies will result from the instrumental resolution, but not enough to account for the observed data.  The Hamiltonian in equation \ref{eq:Hamiltonian_2} will not give two modes that are separated by $\Delta E \sim 10$ meV at the Y point without severely compromising the energies at other points in the Brillouin zone.  These discrepancies hint at the need to allow for effects beyond the theory used here, or for extra terms in the Hamiltonian.

\section{Discussion}

While there are limitations in the ability of the analysis to describe the measured neutron intensity, linear spin wave theory using equation \ref{eq:Hamiltonian_2} and the parameters listed in table \ref{tab:Exchanges} allow for significant insight into the spin dynamics of NiPS$_3$.

The strength of the interplanar exchange, $J^{\prime}$ is particularly noteworthy.  It is the strongest of all the transition metal-PS$_3$ determined to date, amounting to $\sim$10\% the value of $J_1$, hence NiPS$_3$ is more three-dimensional than its sister compounds.  This helps explain the magnitude of the N{\'e}el temperature which, at $T_N \sim 155$ K, is considerably larger than MnPS$_3$ (78 K), FePS$_3$ (123 K) and CoPS$_3$ (120 K), especially when combined with the very strong antiferromagnetic $J_3$ which, again, is much stronger than any of the exchanges determined in the sister compounds.  It also explains the observation from Raman spectroscopy that NiPS$_3$ maintains a long-ranged magnetic structure when thinned down to two monolayers, but does not order when thinned to a monolayer \cite{Kim}.  Without a relatively strong interplanar exchange, the magnetic anisotropy is insufficient to stabilise long-ranged order.

The relatively weak anisotropy is predominantly XY-like, which was also a conclusion from Raman spectroscopy measurements \cite{Kim}, however a small uniaxial anisotropy must be present to force the moments to align almost parallel to the {\bf{a}} axis.  The analysis allows for this anisotropy through the inclusion of $D^x$.  The fitted magnitude may be compared with observations of a spin-flop transition in NiPS$_3$ when a field is applied in the \emph{ab} planes, given as $B_{SF} \approx 6$ Tesla and $B_{SF} \approx 14$ Tesla based on magnetometry \cite{Basnet} and $^{31}$P NMR \cite{Bougamha} respectively.  Mean field theory gives the spin flop transition at \cite{Blundell}:
\begin{equation}
  \label{eq:SpinFlop}
  B_{SF} = \sqrt{\frac{D^x}{M^2}\left(2AM^2 - D^x\right)},
 \end{equation}
where $M$ is the ordered moment, which is is 1.05 $\mu_B$ from neutron diffraction \cite{Wildes15},  and
\begin{equation}
  \label{eq:Afac}
  A=-\frac{4zJ_{eff}}{N\left(g\mu_B\right)^2}
 \end{equation}
 where $z$ is the number of nearest neighbours and $J_{eff}$ is an effective nearest-neighbour exchange.   The fitted parameters in table \ref{tab:Exchanges} give $D^x = -0.01$ meV and a weighted mean of $J_{eff} = 2.825$ meV.  Using these with $z = 3$ for the honeycomb lattice gives a spin flop field of $B_{SF} = 7.3$ Tesla, which is close to the values determined from experiment and giving confidence that $D^x$ is a believable value.

As a final point, the quantitative nature of neutron scattering for studying lattice and magnetic dynamics must be reinforced.  Neutron spectroscopy allows access to dynamics over the whole Brillouin zone.  The first Born approximation holds very well for neutron spectroscopy, thus the cross-section is proportional to the dynamic structure factor $S\left({\bf{Q}},\Delta E\right)$ and the measured intensity distribution may be directly compared with calculations from first principles.  The parameters presented in table \ref{tab:Exchanges} represent the best description of the magnetic dynamics to date, however the discrepancies between the measured and calculated intensities for the low-energy modes show that elements are lacking in the relatively straight-forward analysis and further work is needed.  Any future discussion, though, must be able to produce quantitatively the spin wave energies and neutron intensities apparent in the data presented here.



\section{Conclusions}
Neutron spectroscopy has been used to measure the one-magnon spin dynamics of NiPS$_3$.  The data have been analysed with linear spin wave theory using a Heisenberg Hamiltonian with two single-ion anisotropies.  The analysis successfully reproduces the measured spin wave energies, and the resulting exchange parameters and anisotropies represent the best estimates to date for reproducing the dynamical structure factor.  The analysis does not quite reproduce the measured neutron intensities, however, particularly for the spin fluctuations at the lowest energies, hinting at other more exotic dynamics in the compound.

\begin{acknowledgments}
We thank the Institut Laue-Langevin, experiment number 4-01-1608 \cite{Wildes_IN8_Jun19b}, and Australian Nuclear Science and Technology Organisation (ANSTO), experiment number DB 13377, for the allocation of neutron beam time.  Experiments at the ISIS Neutron and Muon Source were supported by a beamtime allocation RB2010329 \cite{Wildes_MERLIN_Oct20} from the Science and Technology Facilities Council.  We also thank Dr. A. Piovano and Dr. A. Ivanov, F. Charpenay and V. Gaignon for assistance with the IN8 experiment, and J. Shalala, the sample environment group, and Dr. A. Stampfl at ANSTO for assistance with the Taipan experiment.  ARW and KA thank Dr. L. Testa and Prof. H. R{\o}nnow for assistance with the preliminary identification of suitable single crystals using x-ray Laue diffraction. 
\end{acknowledgments}

\appendix
\section{\label{app:evals} Eigenvalues for high-symmetry points}
\begin{equation}
 \label{eq:U}
  \begin{split}
 U =& \left(D^x-D^z\right)\left(D^x+J_1-4J_2-3J_3+2J^{\prime}\right) \\
       & +D^x\left(D^x-D^z+J_1-4J_2-3J_3+2J^{\prime}\right) \\
       & + 2J_1^2 + 8J_2^2 + 4J_3^2 + 4J^{\prime 2} \\
       & - 4J_1J_2 - 4J_1J_3 + 6J_1J^{\prime} \\
       & +12J_2J_3  - 8J_2J^{\prime} - 6J_3J^{\prime}
  \end{split}
\end{equation}

\begin{equation}
 \label{eq:V}
  \begin{split}
 V =& \left(4D^x\right)^2\left(J_1 + J^{\prime}\right)^2  \\
       &+ \left(D^z\right)^2\left(\left(J_1 + J_3\right)^2 + 4\left(J_1 + J^{\prime}\right)^2\right) \\
       &- 16D^xD^z\left(J_1 + J^{\prime}\right)^2 \\ 
       &+ 8\left(2D^x - D^z\right)\left(J_1 + J^{\prime}\right)^2\left(J_1 - 4J_2 - 3J_3 + 2J^{\prime}\right) \\
       &- 16( 2J_1^3J_2 + 2J_1^3J_3 - J_1^3J^{\prime} - 4J_1^2J_2^2 - 6J_1^2J_2J_3\\
       &  + 8J_1^2J_2J^{\prime} - 2J_1^2J_3^2 + 7J_1^2J_3J^{\prime} - 3J_1^2J^{\prime 2}\\
       &  - 8J_1J_2^2J^{\prime} - 12J_1J_2J_3J^{\prime} + 10J_1J_2J^{\prime 2} - 4J_1J_3^2J^{\prime}\\
       &  + 8J_1J_3J^{\prime 2} - 3J_1J^{\prime 3} - 4J_2^2J^{\prime 2} - 6J_2J_3J^{\prime 2} \\
       & + 4J_2J^{\prime 3} - 2J_3^2J^{\prime 2} + 3J_3J^{\prime 3} - J^{\prime 4})
   \end{split}
\end{equation}

\clearpage

\begin{table}
\caption{Table showing the analytic expressions for the eigenvalues at specific points in the Brillouin zone, as given by SpinW.  The parameters $D^x$ and $D^z$ are taken to be always negative, for uniaxial anisotropy, and positive, for planar anisotropy, respectively.  The parameters $U$ and $V$ for the eigenvalues at the Z point are defined in equations \ref{eq:U} and \ref{eq:V}}
\label{tab:Eigenvalues}
\begin{ruledtabular}
\begin{tabular}{ll}
$\Gamma$ & 0 0 0\\
1: & $2S\left(D^x\left(D^x-D^z-J_1-4J_2-3J_3\right)\right)^{\frac{1}{2}}$ \\
2: & $2S\left(\left(D^x-D^z\right)\left(D^x-J_1-4J_2-3J_3\right)\right)^{\frac{1}{2}}$ \\ 
3: & $2S\left(\left(D^x-D^z+2J_1-4J_2+2J^\prime\right)\left(D^x+J_1-3J_3+2J^\prime\right)\right)^{\frac{1}{2}}$ \\
4: & $2S\left(\left(D^x+2J_1-4J_2+2J^\prime\right)\left(D^x-D^z+J_1-3J_3+2J^\prime\right)\right)^{\frac{1}{2}}$ \\
\hline
Y & $\frac{1}{2} 0 0$ \\
1: & $2S\left(\left(D^x-J_3+J^\prime\right)\left(D^x-D^z+J_1-2J_3+J^\prime\right)\right)^{\frac{1}{2}}$ \\
2: & $2S\left(\left(D^x-D^z-J_3+J^\prime\right)\left(D^x+J_1-2J_3+J^\prime\right)\right)^{\frac{1}{2}}$ \\
\hline
Z & $0 \frac{1}{2} 0$ \\
1: & $S\left(2U - 2\left(V\right)^{\frac{1}{2}}\right)^{\frac{1}{2}}$ \\
2: & $S\left(2U + 2\left(V\right)^{\frac{1}{2}}\right)^{\frac{1}{2}}$ \\\\
\hline
C & $\frac{1}{2} \frac{1}{2} 0$ \\
1: & $2S\left(\left(D^x+J^\prime\right)\left(D^x-D^z+J_1-3J_3+J^\prime\right)\right)^{\frac{1}{2}}$ \\
2: & $2S\left(\left(D^x-D^z+J^\prime\right)\left(D^x+J_1-3J_3+J^\prime\right)\right)^{\frac{1}{2}}$ \\
\hline
B & $0 0 {\frac{1}{2}}$ \\
1: & $2S\left(\left(D^x-D^z+4J^\prime\right)\left(D^x-J_1-4J_2-3J_3+4J^\prime\right)\right)^{\frac{1}{2}}$ \\
2: & $2S\left(\left(D^x+4J^\prime\right)\left(D^z-D^x-J_1-4J_2-3J_3+4J^\prime\right)\right)^{\frac{1}{2}}$\\
3: & $2S\left(\left(D^x-D^z+2J_1-4J_2+2J^\prime\right)\left(D^x+J_1-3J_3+2J^\prime\right)\right)^{\frac{1}{2}}$\\
4: & $2S\left(\left(D^x+2J_1-4J_2+2J^\prime\right)\left(D^x-D^z+J_1-3J_3+2J^\prime\right)\right)^{\frac{1}{2}}$\\
  \end{tabular}
\end{ruledtabular}
\end{table}



\bibliography{MPS3}

\end{document}